\documentclass[11pt,reqno]{amsart}
\usepackage{amssymb}
\usepackage{hyperref}
\usepackage{graphicx}
\usepackage{hyperref}
\usepackage{url}
\usepackage{amsfonts}
\usepackage{amsthm}
\usepackage{mathrsfs}
\usepackage{latexsym}
\usepackage{amsmath}
\usepackage{setspace}
\usepackage{pstricks}
\usepackage[latin1]{inputenc}
\usepackage[normal]{subfigure}

\title{Real-time Tool for Affine Transformations of Two Dimensional IFS Fractals}

\author{Elena Hadzieva}
\address{University "St. Paul the Apostle", Partizanska bb, Ohrid, Macedonia}
\email{elena.hadzieva@uist.edu.mk}

\author{Marija Shuminoska}
\address{University "St. Paul the Apostle", Partizanska bb, Ohrid, Macedonia}
\email{marija.shuminoska@cse.uist.edu.mk}

\begin{document}

\begin{abstract}
This work introduces a novel tool  for interactive, real-time transformations of two dimensional IFS fractals. We assign barycentric coordinates (relative to an arbitrary affine basis of $\mathbb{R}^2$) to the points that constitute the image of a fractal.  The tool uses some of the nice properties of the barycentric coordinates, enabling any affine transformation of the basis, done by click-and-drag, to be immediately followed by the same affine transformation of the IFS fractal attractor. In order to have a better control over the fractal, as affine basis we use a kind of minimal simplex  that contains the attractor. We give theoretical grounds of the tool and then the software application.

\end{abstract}


\maketitle

\section{Introduction}

One of the simplest ways of creating fractals is by means of {\it iterated function systems} (IFSs). Generally, the IFS can be defined on any complete metric space, but for the purposes of this paper we will restrict our discussion to the two-dimensional real space $ \mathbb{R}^2$. In this section we will go through the basic definitions, given in \cite{Barnsley, Falconer1}. The transformation defined on  $ \mathbb{R}^2$ of the form  $$w(\mathbf{x})=\mathbf{A} \cdot\mathbf{x}+\mathbf{b}, $$ where $\mathbf{A}$ is $ 2\times 2$ real matrix and $ \mathbf{b}$ is  a two-dimensional real vector,  is called a (two-dimensional) {\it affine transformation}. The finite set of affine contractive transformations, $ \{w_1, w_2, \ldots, w_n\}$ with respective contractivity factors $s_i, \,\, i=1,2,\ldots, n,$ together with the Euclidean space $ (\mathbb{R}^2, d_E)$ is called {\it (hyperbolyc) iterated function system (IFS)}. Its notation is $ \{\mathbb{R}^2; w_1, w_2, \ldots, w_n\}$ and its {\it contractivity factor} is $ s=\max\{s_i, i=1,2, \ldots, n\}$. Since  $(\mathbb{R}^2, d_E)$ is a complete metric space, such is $(\mathcal{H}(\mathbb{R}^2), h(d_E))$, where $(\mathcal{H}(\mathbb{R}^2), h(d_E))$ is the space of nonempty compact subsets of $\mathbb{R}^2$, with the Hausdorff metric $ h(d_E)$ derived from the Euclidean metric $d_E$ (\cite{Barnsley}). It can be shown that if the hyperbolic IFS $ \{\mathbb{R}^2; w_1, w_2, \ldots, w_n\}$ with the contractivity factor $s$ is given, then the  transformation $ W$ defined on $ \mathcal{H}(\mathbb{R}^2)$, by
$$ W(B) = \cup_{i=1}^nw_i(B), \qquad B\in  \mathcal{H}(\mathbb{R}^2$$
is a contraction mapping with the contractivity factor $s$. According to the fixed-point theorem, there exist a unique fixed point of $W$, $ A\in  \mathcal{H}(\mathbb{R}^2)$, called the {\it attractor} of the  IFS, that obeys $W(A)=A=\lim_{n\rightarrow \infty} W(B)$, for any $ B\in  \mathcal{H}(\mathbb{R}^2)$. (Note that there are weaker conditions under which the attractor of an IFS can be obtained - the IFS need not to be contractive, see \cite{Barnsley1}.)

Two algorithms are used for the visualization of an IFS attractor, deterministic and random (the last is also called chaos game algorithm). In this paper we construct the images of the attractors with the second,  more efficient algorithm.  Our primary target in this paper will be fractal attractors, since  there is no  interactive, real-time tool for their modeling, to the best of our knowledge..   

The rest of the paper is organized in four sections: Related Articles, Theoretical Grounds of the Tool, The Tool and its Application, Conclusions and Future Work.

\section{Related Articles}

Barnsley et al., in   \cite{Barnsley2, Barnsley3} define the fractal transformation as mapping from one attractor to another, using the ``top" addresses of the points of both attractors and  illustrate the application in digital imaging. Although not user friendly, not real-time and not continuous (they are continuous under certain conditions), these transformations, have a lot of potential for diverse applications. In \cite{Chang} the coordinates of the points that compose the fractal image are determined from the IFS code and then the IFS code is modified to obtain translated, rotated, scaled or sheared fractal attractor, or attractor transformed by any affine transformation as the composition of aforementioned transformations. To make a new transformation, another IFS code has to be constructed. In \cite{Darmanto}, Darmanto et al. show a method for making weaving effects on tree-like fractals. They make a local control of the ``branches" of the tree, by changing a part of the IFS code. The control over the change of the attractor depends on the experience of the user for predicting how the modification of the coefficients in the IFS code will change the tree-like fractal. Compared to the methods for transforming fractals in the cited papers, our tool is user-friendly, real time, relatively fast and relatively low memory consuming. It enables a continuous affine transformation of an arbitrary IFS attractor.

The idea for modeling fractals, i.e. their predictive, continuous transformation, where barycentric coordinates are involved, is exposed in \cite{Kocic2},\cite{KoSi2}, \cite{doktorska}, by means of, so-called {\it Affine Invariant Iterated Function Systems - AIFS}. The IFS code is transformed into AIFS code which involves barycentric coordinates.  The method that we propose is also based on barycentric coordinates, but no transformation of the IFS code is needed. 

Also, Kocic et al. in \cite{KoStBa1} propose  the use of, so called {\it minimal canonical simplex},  for better control of the attractor (for 2D case, minimal canonical simplex is the isosceles right-angled triangle with the minimal area that contains the attractor and whose catheti are parallel to the {\it x} and {\it y} axes). In \cite{BaKo1} the authors prove a theorem for existence and uniqueness of such minimal canonical simplex. It is the limit of the Cauchy sequence of the minimal simplexes of the $n$-th preattractor  when $n$ approaches to infinity, in the complete metric space  $(\mathcal{H}(\mathbb{R}^m), h(d_E))$. (The $n$-th preattractor in this paper  is defined as the set obtained after $n$ iterations of the chaos game algorithm.)

\section{Theoretical Grounds of the Tool}
Barycentric coordinates of points (and vectors) promote global geometrical form and position rather than exact coordinates in relation to the origin/axes.  Thus, the geometric design done by the use of barycentric coordinates can be called {\it coordinate-free geometric design}.

%\footnote{The idea for this term originated from the subtitle of the manuscript \cite{Derose},  "Coordinate-free Approach".} 

The set of the three points $ \{A, B, C\}$ is said to be {\it affine basis} of the affine space of points $\mathbb{R}^2$ if the set $ \{\overrightarrow{AB}, \overrightarrow{AC}\}$ is a vector base of $\mathbb{R}^2$, observed as a vector space (\cite{Gallier}). We say that the point $M$ has {\it barycentric coordinates} $ (a, b, c)$ {\it relative to the basis} $ \{A, B, C\}$, where $ a+b+c =1$, and we write 
\begin{equation}\label{osnovna} M= a A+bB+ cC \end{equation} 
if and only if one of the following three equivalent conditions holds:
\begin{eqnarray*}
\overrightarrow{AM} &=& b\overrightarrow{AB} +c\overrightarrow{AC}\\
\overrightarrow{BM} & =&  a\overrightarrow{BA} + c\overrightarrow{BC}\\
\overrightarrow{CM} &=& a\overrightarrow{CA} + b \overrightarrow{CB}
\end{eqnarray*}

We will give the relations between the barycentric coordinates relative to the given basis $ \{A, B, C\}$ and the rectangular coordinates of an arbitrary point from $ \mathbb{R}^2$. Suppose that the arbitrary point $M$ and the three basis points $ A, B$ and $C$ have rectangular coordinates $ (x, y)$, $(a_1, a_2)$, $(b_1, b_2)$ and $(c_1, c_2)$, respectively. Then the relation (\ref{osnovna}) can be rewritten in the following form \begin{equation} \left[\begin{array}{c}x\\y\end{array}\right] = a  \left[\begin{array}{c}a_1\\a_2\end{array}\right] + b \left[\begin{array}{c}b_1\\b_2\end{array}\right] +c  \left[\begin{array}{c}c_1\\c_2\end{array}\right],\end{equation} or, in matrix form $$ \left[\begin{array}{c}x\\y\end{array}\right] = \left[\begin{array}{ccc}a_1&b_1&c_1\\a_2&b_2&c_2\end{array}\right]\left[\begin{array}{c}a\\b\\c\end{array}\right].$$

By rearranging the last matrix equation, we obtain more suitable form for further manipulations:
 \begin{equation} \label{bar-vo-dek}\left[\begin{array}{c}x\\y\\1\end{array}\right] = \left[\begin{array}{ccc}a_1&b_1&c_1\\a_2&b_2&c_2\\1&1&1\end{array}\right]\left[\begin{array}{c}a\\b\\c\end{array}\right].\end{equation}

The relation (\ref{bar-vo-dek}) defines the {\it conversion of the barycentric coordinates relative to the affine basis $ \{A, B, C\}$ into the rectangular coordinates}. Let us denote the $ 3\times 3$ matrix by $\mathbf{T} $ to indicate ``triangle". The inverse conversion can be easily get, by simple multiplication of (\ref{bar-vo-dek}) by the inverse of $\mathbf{T} $ (which exists, since $ \{A, B, C\}$ is an affine basis). That is, the {\it conversion of the rectangular coordinates into the barycentric coordinates relative to the basis $\{A, B, C\}$} is defined by the relation
\begin{equation} 
\left[\begin{array}{c}a\\b\\c\end{array}\right] = \mathbf{T}^{-1} \cdot \left[\begin{array}{c}x\\y\\1\end{array}\right],
\end{equation} 
which after calculating the inverse of $\mathbf{T}$, can be expressed in its explicit form,

\begin{equation} \label{dek-vo-bar} \left[\begin{array}{c}a\\b\\c\end{array}\right] = \frac{1}{\det {\mathbf T}}\left[\begin{array}{ccc}b_2-c_2&c_1-b_1&b_1c_2-b_2c_1\\c_2-a_2&a_1-c_1&a_2c_1-a_1c_2\\a_2-b_2&b_1-a_1&a_1b_2-a_2b_1\end{array}\right] \cdot \left[\begin{array}{c}x\\y\\1\end{array}\right],
\end{equation}
where $ \det {\mathbf T} = a_1b_2-b_1a_2+b_1c_2-c_1b_2+c_1a_2-a_1c_2 \not= 0$.

Barycentric coordinates have two other names: {\it affine} and {\it relative} coordinates. Affine, since they are related to an affine basis, and relative, since they define relative positions in the plane/space. Namely, if we consider a set of points with given barycentric coordinates relative to the basis $\{A, B, C\}$ and we affinly transform the basis into the basis $ \{A', B', C'\}$, then the set of points having the same barycentric coordinates, but now relative to the basis $ \{A', B', C'\}$, will keep the relative geometry, i.e. will be transformed by the same affine transformation. Note that any change of one affine basis to another, defines a unique affine transformation. 

Since the image of the attractor, generated by the random algorithm is a finite set of points, we will use the aforementioned property of the barycentric coordinates to affinly transform the IFS fractal.

%Namely,  if $(a, b, c)$ are barycentric  coordinates of the point $M$, relative to the basis $ \{A, B, C\}$, and one of the points, let's say $ A$, is changed to the point $ A'$, then the ratio of the distances of the point $M'$, whose barycentric coordinates relative to the basis $ \{A', B, C\}$ are $ (a, b, c)$, to the points  $ A', B, C$ will be the same as the ratio of the distances of $M$ to the points $ A, B, C.$ (Proof???/affine invariance of the coordinates of a point when affine transformation of the triangle occurs???) The same conclusion can be drawn for any affine transformation of the basis. (Any change of the affine basis $ A, B, C$ into another affine basis defines a unique affine transformation.)

%Since the image of the attractor, generated by the "random algorithm" is a finite set of points all those points will keep the ratios of the distances to the points of the affine basis before and after the affine transformation of the affine basis. Therefore, the whole attractor will be affinly transformed (Proof???).  

\section{The Tool and Its Application}

Our real time, user-friendly tool is written in C$\#$ programming language using the Visual Studio 2013. It allows both on-click definition of the triangle and definition by specifying the triangle's vertices directly in the code, with their rectangular coordinates. When the affine base is defined, the image of the fractal is created (relative to the triangle) and ready for transformation.  

Triangle's vertices are moved by drag and drop option of the cursor.  When the cursor is placed over the vertices, it has a ``hand'' shape, otherwise its shape is ``cross''. The affine transformations of the triangle, immediately followed by the same affine transformation of the fractal image, are visible in real time, in the coordinate system of the window. 

We will show the superiority of our tool on two examples of fractal attractors, the so called ``flower" and ``maple".  In order to get clearer image of the attractor, we neglect the first 14 points and start to plot from the 15-th point.

Note that, by default, the origin of the coordinate system of the window is located at the top-left corner of the window, with positive directions: to the right for the $x$-axis and down for the $y$-axis. 

%We neglect the starting 20 iterating points and start to plot the points from 21-st point in order to get clearer image of the attractor (????).

%The tool is written in Visual studio, in C$\#$. It is user-friendly, it allows defining the triangle (affine basis) "on click", but it can be also defined by specifying the affine basis in the code, with the Descartes coordinates. It enables real time affine transformation of fractals. Two examples will follow.

Example 1. The IFS $ \{\mathbb{R}^2; w_1, w_2\}$, whose attractor is  the ``flower" fractal, is defined by the following contractive mappings $w_i:
\mathbf{x}\mapsto \mathbf{A}_i\mathbf{x}+\mathbf{b}_i$, $ i=1, 2$: 
$$\mathbf{A}_1=\left[\begin{array}{cc} 0.47& 0.30\\-0.30&0.47
\end{array} \right], \,\,\,  \mathbf{b}_1= \left[ \begin{array}{c} 0.37\\ 1.74 \end{array}\right];$$ $$
\mathbf{A}_2=\left[\begin{array}{cc} 0.48& -0.29\\0.29& 0.48
\end{array} \right],\,\,\,  \mathbf{b}_2= \left[ \begin{array}{c}-0.34\\ 1.75 \end{array}\right].$$

Figure \ref{Flower} (a), b), c)) depicts that the arbitrary control triangle is defined on-click. When moving its vertexes by click-and-drag, the fractal attractor appropriately responds to the transformations done over the triangle.

Example 2. The attractor of the IFS $ \{\mathbb{R}^2; w_1, w_2, w_3, w_4\}$, where the mappings $w_i:
\mathbf{x}\mapsto \mathbf{A}_i\mathbf{x}+\mathbf{b}_i$, $ i=1,2,3,4,$ are defined by
$$\mathbf{A}_1=\left[\begin{array}{cc} -0.04& 0\\-0.23& -0.65
\end{array} \right], \,\,\,  \mathbf{b}_1= \left[ \begin{array}{c} -0.08\\ 0.26 \end{array}\right];$$ $$
\mathbf{A}_2=\left[\begin{array}{cc} 0.61& 0\\0& 0.31
\end{array} \right],\,\,\,  \mathbf{b}_2= \left[ \begin{array}{c}0.07\\ 3.5 \end{array}\right];$$
$$\mathbf{A}_3=\left[\begin{array}{cc} 0.65& 0.29\\-0.3& 0.48
\end{array} \right],\,\,\, \mathbf{b}_3= \left[ \begin{array}{c}0.74\\ 0.39
\end{array}\right]; $$ $$\mathbf{A}_4=\left[\begin{array}{cc} 0.64& -0.3\\0.16& 0.56
\end{array} \right],\,\,\, \mathbf{b}_4= \left[ \begin{array}{c}-0.56\\ 0.60 \end{array}\right];$$ is the fractal called ``maple".

We  computed the coordinates of the vertices of the minimal canonical simplex for this fractal and used the vertices as an affine basis.  Such affine basis ensures better control over the attractor (\cite{KoStBa1}, \cite{BaKo1}), see Figure~\ref{Maple}. We examined creation of the shadow or wind effect for the maple tree, throughout d) to f) on this figure.

\section{Conclusions and future work}

We succeeded  in creating a real-time and user-friendly tool for interactive modeling of IFS attractors, with focus on fractal attractors. To the best of our knowledge, this is the first real-time interactive tool for modeling fractals. The tool is very efficient because it is relatively low memory and time consuming. We foresee a great application of our tool, since all images coded by an IFS code can be affinly transformed according the user needs, only by click-and-drag. 

These are the first steps in fractal modeling with our tool. There is much more to be done to enhance the tool's performances. For instance, including a control polygon of more than three points. Moreover, it would be very efficient to use the convex hull of the attractor as a control polygon. In the last case the control over the global changes of the attractor will be maximal. Including local control would be also useful for satisfying the sophisticated needs of the users. Finally, expanding the tool over 3D IFS attractors will bring higher practical importance to the tool. 

%very interesting and  important is extension of  our tool over 3D attractors.

\end{document}